\documentclass[%
 reprint, amsmath,amssymb, aps]{revtex4-2}
\usepackage{graphicx}
\usepackage{dcolumn}
\usepackage{bm}
\usepackage{physics}

\begin{document}

\preprint{ARXIV/2112.15503}

\title{On the strangeness of quantum mechanics}

\author{Marcello Poletti}
 \email{epomops@gmail.com}
\affiliation{%
 San Giovanni Bianco, Italy
}%
\date{\today}

\begin{abstract}
The extravagances of quantum mechanics (QM) never fail to enrich daily the debate around natural philosophy. Entanglement, non-locality, collapse, many worlds, many minds, and subjectivism have challenged generations of thinkers. Its approach can perhaps be placed in the stream of quantum logic, in which the “strangeness” of quantum mechanics is “measured” through the violation of Bell’s inequalities and, from there, attempts an interpretative path that preserves realism yet ends up overturning it, restating the fundamental mechanisms of QM as a logical necessity for a strong realism.
\end{abstract}

\maketitle


\section{Introduction}
Quantum mechanics is strange and has been since its origins, Heisenberg immediately realized that he was working on something deeply anomalous, and he wrote to his friend Pauli as early as 1925: \begin{quote}"Everything is still very vague and unclear to me, but it seems that electrons no longer move in orbits\cite{B1}".\end{quote}

Einstein, Podolsky, and Rosen then set a milestone with their celebrated 1935 work\cite{B2} , opening a more stringent and quantitative debate around the nature of this strangeness.

This debate would find a new milestone with Bell in 1964\cite{B3}. With his inequalities, Bell set down a precise measurable limit between what is ordinary and what is not ordinary.
The Bell inequalities measure the strangeness of quantum mechanics and, subject to experimental verification, the strangeness of the world.

Perhaps the most dry and simple version of Bell’s inequality\cite{B14} is the Wigner-D’Espagnat\cite{B4} (WE) inequality, which we restate in the following form:
\begin{equation*}
	\{A,B\}+\{\neg B,C\}\geq\{A,C\}
\end{equation*}
"The things that are \textbf{A and B} united with the things that are \textbf{not B and C} include things that are \textbf{A and C}".

In fact, if \textbf{x} is in $\{A,C\}$  then it will be \textbf{B} or \textbf{not B}. In the first case \textbf{x} is part of $\{A,B\}$ in the second of $\{\neg B,C\}$. 
From this elementary property of sets follows a counterpart related to probabilities:
\begin{equation*}
	P(A,B)+P(\neg B,C)\geq P(A,C)
\end{equation*}
Quantum mechanics violates this inequality, for example, in a Stern-Gerlach experiment with magnets oriented at 0° and 45°. 
Indicating with $N_\alpha$  (risp. $S_\alpha$) the event "The electron goes North (risp. South) in a field with orientation $\alpha$" we have the inequality:
\begin{equation*}
	P(N_0,S_{45})+P(N_{45},S_{90})\geq P(N_0,S_{90})
\end{equation*}
On the other hand, the following formula applies in quantum mechanics:
\begin{equation*}
	P(N_\alpha,S_\beta)=\frac{1}{2}sin^2(\frac{\beta-\alpha}{2})
\end{equation*}
Executing the calculations:
\begin{equation*}\begin{split}
		P(N_0,S_{45})+P(N_{45},S_{90})\approx 0.15 \\
		P(N_0,S_{90})\approx 0.25
\end{split}\end{equation*}
The inequality is violated by quantum mechanics, which is therefore strange to the extent that it violates something structural in WE’s elementary proof.

Even more than theory, of course, experimental verification, particularly from Aspect\cite{B5} onward, poses an even more stringent problem. The world is strange, regardless of whether quantum mechanics is correct or not.
\section{Tertium non datur}
Note that WE’s proof exploits the principle of the excluded third in the form "if \textbf{x} is in $\{A,C\}$  then it will be \textbf{B} or \textbf{not B}" [*].
The problem that arises is that in reality [*] is very weak if the property B, referring to x, is undecidable, where this concept of undecidability is independent of the incompleteness theorems and can be carried over to elementary logic in a completely trivial form.

Consider the premises:
\begin{itemize}
	\item Socrates is a man.
	\item All men are mortal.
\end{itemize}
And the conclusion
\begin{itemize}
	\item \textbf{p}: Socrates has a beard
\end{itemize}
All that matters here is that the conclusion \textbf{p} is undecidable with respect to the premises in the sense that neither the truth nor the falsity of \textbf{p} can be deduced from the premises.

More generally, given \{S\} a system of postulates (axioms, premises), we say that \textbf{p} is undecidable in \{S\} if, by adding \textbf{p} or \textbf{not p} to \{S\}, a contradictory system is not produced.
By including undecidable propositions (or properties), it is possible to falsify WE simply by considering that it \{A,C\} can contain elements \textbf{x} that are not \textbf{B} and are not \textbf{not B}.

In this sense WE indicates a limit to the scope of properties considered: it is true if the scope is that of only decidable properties (propositions), and is to be rejected if it is not.
To include undecidable propositions within a coherent algebra, since it is not possible to assign to these propositions the classical values \{0, 1\}, we enter the field of multi-valued logic, of which there are several definitions\cite{B6}, and cascading into the general field of fuzzy sets.

Multi-valued logics and fuzzy sets produce various probability theories, of which QM is clearly a particularly effective example. In fact, barring the dynamic postulate [Schrödinger equation], QM is nothing more than a particularly sophisticated probability-calculating tool. We note, however, that the level of sophistication is not essential in the sense that even the simplest standard fuzzy logic\cite{B7}, while proving inadequate, violates WE. [Appendix \ref{Appendix}]. 

What is proposed here is to ground such algebras on the concept of undecidability or, equivalently, to base the probability on that of information, understood as "implicit deductive capacity".

In this sense we propose to legitimize, with respect to the given axioms, a proposition like "The probability that Socrates has a beard is 0.5" as a measure of the most absolute impossibility of arguing for or against the thesis within the scheme of the given premises.

Moreover, we expect such an algebra to be structurally complex, since it cannot be based on syntactic analysis alone, but inevitably on semantic analysis as well. Consider the following premise:
\begin{itemize}
	\item A box contains a yellow, red, or green apple.
\end{itemize}
And the undecidable proposition:
\begin{itemize}
	\item \textbf{p}: The apple in the box is red.
\end{itemize}
There is a subtle semantic connection between the two propositions that leads us to consider assigning p a truth or expectation value of 1/3. This relationship emerges most clearly by considering "$p\land\neg p$" or "$p\lor\neg p$" type propositions that are decidable despite the undecidability of p. In this sense, the standard fuzzy logic has been called "inadequate": the expected algebra cannot be mere expectation value algebra but will have to manage the mutual relationships between propositions, for example treating them as vectors variously oriented with respect to each other.

More specifically, it is interesting to estimate the number of degrees of freedom expected for an algebra of this type. Let p and q be two undecidable propositions; imagining a vector algebra in which to place them, p and q exhibit two distinct types of "perpendicularity":
\begin{enumerate}
	\item {	p and q are perpendicular in the sense of incompatibility, like $p$ e $\neg p$}.
	\item {p and q are perpendicular in the sense of independence, like}
	\begin{itemize}
		\item Socrates has a beard.
		\item Aristotle is tall.
	\end{itemize}
\end{enumerate}
Where 1) defines a two-dimensional vector space (Tertium non datur) and 2) a vector space with dimensions at most equal to the number of propositions under consideration.
An algebra that incorporates these relations will therefore not be a real value algebra (as in fuzzy logic) while a vector algebra on a complex field such as QM could be.
\section{From logic to physics}
In 1814, Pierre-Simon de Laplace published the "Essai philosophique sur les probabilités", in whose introduction he described the famous summation of nineteenth-century determinism, which later became known as "Laplace’s demon":
\begin{quote}
	"We may consider the current state of the universe as the effect of its past and the cause of its future. Given an intellect which at a certain point should know all the forces that set nature into motion, and all the positions of all the objects of which nature is composed, if this intellect were also sufficiently expansive as to submit this information to analysis, it would include in a single formula the movements of the largest bodies of the universe and those of the smallest atoms; for such an intellect nothing would be uncertain and the future itself like the past would be evident before its eyes."\cite{B8}
\end{quote}
A stronger assumption also applies in a classical context. A demon that knew with absolute precision even just a finite portion, small as you like, of "true"  Newtonian space\cite{B9} and for a finite period of time, small as you like, i.e., a demon that knew every detail of a space-time sphere as small as you like, could deduce (calculate) the past, present and future of the entire universe. This, in the final analysis, is guaranteed by the analyticity of the classical gravitational field, its properties known within a finite area in space and time, it is possible to deduce its global form by analytical extension. In this sense, classical mechanics is devised such that \textit{all the information about the entire history of the universe is available in any space-time sphere as small as you like}.

If analyticity is broken (suppose, for example, that the force of universal gravitation is instantaneously zero for distances greater than X, with X also very large), Laplace’s demon could still calculate past, present, and future while the strongest assumption would no longer apply; for a “local demon” there would be properties of the universe that would be inaccessible, and we would say, by obvious extension, undecidable.

Notice that undecidable properties were obtained in a completely deterministic system; the determinism-indeterminism dialectic is not central here. What really matters is the inaccessibility of part of the information within a local system. Unlike the classical system, in the system with non-analytical correction, what happens is that the information is no longer fully available everywhere but, in general, a given system A, within a given time interval, has access to only a fragment of that information.
From this parallel with logical systems emerges, for each observer O local information (Axioms) is available, deterministically connected to remote properties not in O (Theorems), but there are also undecidable properties available that, with respect to O, will not be "entangled" with O.

These properties are indeterministic in O, but in a sense that has nothing to do with deterministic chaos and not even with the intuitive idea of randomness as the result of a random process that generates unpredictable values; here we are dealing with indeterminism triggered by the most profound lack of information.
The construction of a scientific theory capable of algebraically handling the expectation values of undecidable properties must be based on algebras of the type discussed in the second chapter, QM being a concrete and effective example of such a theory.

A physical system can thus violate WE, and Bell’s inequalities now mark a boundary between local determinism (the strong version of Laplace’s demon) and its negation.
\section{Relativism and Information}
In a universe where local determinism is denied, a system A will describe a system B with algebras that can in general violate Bell’s inequalities. Clearly this is a relative situation; B in its turn will describe A in the same way.
The emergence of quantum states occurs through tearing of causal connections and subsequent increase in undecidable properties. Essentially it is isolation, understood as the separation of information, which triggers the need in A to describe B through tools such as QM.

From this point of view, the essence of Schrodinger’s  cat experiment\cite{B10} is all in the box, which separates the two systems, as long as this box ensures a profound impossibility of knowing its content; such content can only be described through a mathematics capable of violating Bell’s inequalities; conversely - and this is point is crucial - the cat will be in the position to describe the experimenter outside of the box through something analogous to a quantum superposition.

Consider this mental experiment.
Suppose you have two boxes. In box A there is Alice, in box B there is Bob, each with a coin. Both use their own technology to isolate the other until they are in a genuine quantum state.
In this symmetrical situation it is legitimate, if not necessary, to invoke Einstein’s  principle of relativity\cite{B11}: the form of the laws of physics must be identical in all reference frames. The interpretation of this principle in our case assumes the following form:
If Alice describes Bob’s coin with a quantum state $\alpha\ket{Head}+\beta\ket{Tail}$ then Bob must do the same with Alice’s coin.

The analysis conducted thus far therefore leads in the direction of relational interpretation\cite{B12}, but devoid of the last remnants of “plural ontology” present there in the form of “subjective reality” or “relative reality”.

At the risk of hazarding an inappropriate comparison, the proposal under discussion is to make a separation in physics similar to what Gödel\cite{B13} did with the concepts of truth and provability. Within a system \{S\} a property is undecidable and \textbf{therefore} must be treated with tools such as QM without detracting from the fact that the property is objectively “true” or “false” in larger macrosystems.
\section{Philosophy}
Philosophically, the mechanism in classical mechanics by which information is fully available everywhere is wholly compatible with a radical solipsism, since nothing prohibits thinking of the universe as an (analytical) extension of one’s own Cartesian ego. Strictly speaking, the application of Occam’s razor tends to impose such a choice.

On the contrary, a strong realism, which could be summarized in the motto “there is something beyond my Cartesian ego”, should be based on a strong ontology of the sense of “existing”, which cannot fall into radical solipsism, that is, which cannot derive from a simple extension of the self.
In other words, B exists for A precisely to the extent to which B possesses undecidable properties for A.

In this sense QM, or more generally the violation of WE, is a philosophical necessity for those who believe that realism is an important philosophical value. In a flip-flop, classical mechanics, the historical temple of realism, is seen as inadequate for this purpose, and QM, the historical temple of idealism, is seen as a necessity for this purpose.
More precisely, QM is seen as a good candidate for a strong local realism.

It should be entirely obvious from what has been discussed here that the proposed approach does not require any kind of remote interaction but, much more banally, merely the violation of WE in a system separate from others, which, on the contrary, may be in the entangled regime and thus mutually deterministic without invoking any mechanism of action between the two.
\section{Conclusions}
The possibility of thinking of QM as an “algebra of undecidable propositions” has been discussed here.
This leads to the conclusion that the only concrete fact derivable from violating Bell’s inequalities is the observation that in the real world, the assumption of Laplace’s demon in its strongest local version does not hold; each subset A of the universe U contains only part of the information available in U. In short: “The universe is the smallest description of itself”

It is also concluded that the relational interpretation is substantially correct, including in a drier form that foregoes residual subjectivism. QM \textbf{describes} the world from a certain perspective, without needing to assign any ontological weight to this description.
Finally, it was indicated how this approach is compatible with a strong local realism; and moreover, how it emerges as a necessity for not stumbling into the most radical solipsism.

\appendix

\section{Fuzzy logic violates WE}\label{Appendix}

Given T, the chosen t-norm and S, the corresponding s-norm, WE is written as
\begin{equation*}
	T(a,c)\leqslant S(T(a,b),T(1-b,c))
\end{equation*}
Let $b=0.5<a<c$. And let t-norm $T(x,y)=min(x,y)$ be given, hence s-norm $S(x,y)=max(a,b)$. So:

\begin{equation*}
	\begin{split}
		a&=min(a,c) \\
		&>max(min(a,b),min(1-b,c)) \\
		&=max(b,b) \\
		&=b		
	\end{split}
\end{equation*}

\nocite{*}

\bibliography{Eng}

\end{document}